# Observation of unconventional van der Waals multiferroics near room temperature

Yangliu Wu[1], Haipeng Lu[1], Xiaocang Han[2], Chendi Yang[3], Nanshu Liu[5], Xiaoxu Zhao[2], Liang Qiao[4], Wei Ji[5], Renchao Che[3], Longjiang Deng[1]* and Bo Peng[1]*

[1]National Engineering Research Center of Electromagnetic Radiation Control Materials, School of Electronic Science and Engineering, University of Electronic Science and Technology of China, Chengdu 611731, China
[2]School of Materials Science and Engineering, Peking University, Beijing 100871, China
[3]Laboratory of Advanced Materials, Department of Materials Science, Collaborative Innovation Center of Chemistry for Energy Materials(*iChEM*), Fudan University, Shanghai 200433, China
[4]School of Physics, University of Electronic Science and Technology of China, Chengdu 611731, China
[5]Beijing Key Laboratory of Optoelectronic Functional Materials & Micro-Nano Devices, Department of Physics, Renmin University of China, Beijing 100872, China
*To whom correspondence should be addressed. Email address: bo_peng@uestc.edu.cn; denglj@uestc.edu.cn

**Abstract:** The search for two-dimensional (2D) van der Waals (vdW) multiferroics is an exciting yet challenging endeavor. Room-temperature 2D vdW few-layer multiferroic is a much bigger insurmountable obstacle. Here we report the discovery of an unconventional 2D vdW multiferroic with out-of-plane ferroelectric polarization and long-range magnetic orders in trilayer $NiI_2$ device from 10 K to 295 K. The evolutions of magnetic domains with magnetic field, and the evolutions between ferroelectric and antiferroelectric phase have been unambiguously observed. More significantly, we realize a robust mutual control of magnetism and ferroelectricity near room temperature. The magnetic domains are manipulated by a small voltage ranging from ±1 V to ±6 V at 0 T and 295 K. This work opens opportunities for exploring multiferroic physics at the limit of few atomic layers.

Multiferroic materials with a coexistence of ferroelectric and magnetic orders has been diligently sought after for a long time to achieve the mutual control of electric and magnetic properties toward the energy-efficient memory and logic devices[1-3]. But the two contrasting order parameters tend to be mutually exclusive in a single material[4]. Nondisplacive mechanisms introduce a paradigm for constructing multiferroics beyond the traditional limits of mutual obstruction of the ferroelectric and magnetic orders[5,6]. To date, the type I multiferroic $BiFeO_3$ is the only known room-temperature single-phase multiferroic material. Alternatively, the helical magnetic orders break the spatial inversion symmetry and simultaneously lead to electric orders[7,8], giving rise to type-II multiferroics. The quest for a room-temperature single-phase multiferroic remains an open challenge.

The emergence of 2D vdW magnets and ferroelectrics has opened new avenues for exploring low-dimensional physics on magnetoelectric coupling[9,10]. Diverse isolated vdW ferromagnets[11-13] and ferroelectrics[14,15] have enabled tantalizing opportunities to create 2D vdW spintronic devices with unprecedented performances at the limit of single or few atomic layers. Few of bulk crystals of transition-metal dihalides with a trigonal layered structure have been shown that the helical spin textures break inversion symmetries and induce an orthogonal ferroelectric polarization [16,17], but typically with low Néel temperature below 80 K.

A recent work shows debatable evidence of type-II monolayer $NiI_2$ multiferroics using the optical measurements of second-harmonic-generation (SHG) and linear dichrosim (LD)[18]. The main debate is that all-optical characterizations are unreliable to make a judgement of a few- and single-layer multiferroic at the presence of non-collinear and antiferromagnetic orders[19]. A recent work has demonstrated that the helimagnetic structure cannot be persisted in monolayer $NiI_2$[20], and the observed SHG and LD signals in few-layer $NiI_2$ originate from the magnetic-order-induced breaking of spatial-inversion[19,20]. The SHG and LD signals are not necessarily associated with ferroelectric polarization, which only reflect the breaking of spatial inversion symmetry, rather than the ferroelectric polarization. Antiferromagnetic (AFM) and non-collinear magnetic order, ferroelectric polarization and lattice asymmetry all can break spatial inversion symmetry and induce LD and SHG signals[21]. Unfortunately, this is often forgotten and naively the bare existences of optical signals are taken as proof of ferroelectricity, which is a big pitfall for judging the multiferroic of few-layer 2D materials, particularly for single-layer. The prerequisite of ferroelectric polarization is the non-vanishing spontaneous electric polarizations, which must be proven through reliable and direct electrical measurements, such as polarization- and current-electric field (*P-E* and *I-E*) hysteresis loops. To date, 2D vdW multiferroic has not been directly uncovered at the limit of few layers, in particular, a room-temperature 2D vdW multiferroic is a bigger challenge. Here, we report unconventional near-room-temperature vdW multiferroic in trilayer $NiI_2$ device based on magneto-optical-electric joint-measurements, studying the temperature-, frequency-, magnetic-field- and electric-field dependence of magnetic domains and ferroelectric polarization. In this 2D vdW multiferroics, an unprecedented mutual control of magnetism and ferroelectric polarization has been realized.

**Magnetism in trilayer $NiI_2$ device**

Due to the high reactivity of $NiI_2$ flakes, $NiI_2$ exfoliation and encapsulation by graphene and hexagonal boron nitride (hBN) flakes were carried out in a glove box (Fig. 1a and Extended Data Fig. 1). $NiI_2$ crystal shows rhombohedral structure with a repeating stack of three (I-Ni-I) layers, where Ni and I ions form a triangular lattice in each layer (Fig. 1b). The rhombohedral stacking has been atomically identified (Fig. 1c and Extended Data Fig. 2). The atom arrangements of rhombohedral phase demonstrate signature hexagon-shaped periodic bright spots with equal contrast, validating the overlapping stack of I and Ni atoms along the *c* axis. The ADF-STEM and fast Fourier transform (FFT) show an interplanar spacing of 1.9 Å, consistent with the (110) lattice plane of rhombohedral NiI$_2$ crystal. Circularly polarized Raman spectra in the parallel (σ+/σ+ and σ-/σ-) configuration show only two distinct peaks in the NiI$_2$ device (Fig. 1d). The peak at ~124.7 cm$^{-1}$ is assigned to the A$_g$ phonon modes[22], and this polarization behavior is consistent with Raman tensor analysis for the rhombohedral structure of NiI$_2$[23]. The Raman feature at ~20 cm$^{-1}$ is assigned to the interlayer shear mode (SM), which suggests that the NiI$_2$ is trilayer[20].

For optimal optical response and sensitivity to probe the magnetic properties, the photon energy should be chosen near the absorption edge[11,24]. Therefore, we first studied white-light magnetic circular dichroism (MCD) spectra of a trilayer NiI$_2$ device as a function of magnetic field perpendicular to the sample plane at 295 K (see Methods for details)[25]. There is a strong peak near 2.3 eV along with two weak features around 1.85 eV and 1.6 eV (Fig. 1e). By means of ligand-field theory, the peaks are attributed to the absorption transitions of *p-d* exciton states[26]. A pair of opposite MCD peaks with magnetic field manifestly appears at 2.3 eV, suggesting strong magneto-optical resonance. Figure 1f plots corresponding color map of MCD curves as a function of magnetic field. When the magnetic field is switched, MCD features is consistently reversed, and non-zero remanent MCD signal at ~2.3 eV is distinctly observed at 0 T, indicating spontaneous spin-up magnetizations, which induce asymmetric magnetic-field dependence. This suggests that the long-range magnetic orders persist to near room temperature.

To further validate the room-temperature magnetic order, the scanning RMCD microscope was used to image and measure the magnetic domains of the as-exfoliated trilayer NiI$_2$. The polar RMCD imaging is a reliable and powerful tool to unveil the 2D magnetism in the micro scale, and the RMCD intensity is proportional to the out-of-plane magnetization [24]. All magneto-optical measurements were carried out using a 2.33 eV laser with optimal detection sensitivity (see Methods for details). Figure 2a shows room-temperature RMCD maps of a trilayer NiI$_2$ sweeping between -1 T and +1 T. At $\mu_0 H$ = 0 T, the RMCD map shows a complex magnetic texture composed of three zones: magnetization up (yellow and white), magnetization down (sky and dark blue), and no net magnetization (cyan). As the magnetic field is increased to +1 T, the magnetizations in most areas switch to spin-up along with increasing upward magnetization intensities, and the spin-up areas are significantly enlarged. And as the magnetic field is swept down to -1 T, the magnetizations reverse and switch to spin-down, accompanied with increasing spin-down domain area. Although a few spin-up magnetic domains are partly preserved, the spin-up intensity and areas apparently decrease. We have further studied the magnetism of another few-layer NiI$_2$ through RMCD imaging. Similar magnetic controls of magnetic texture are detected (Extended Data Fig. 3). Spin-up magnetizations increase as positive magnetic fields increase to +0.75 T, and switch to spin-down as the magnetic fields reverse to -0.75 T. This observation of magnetic control of magnetic domains is the hallmark of room-temperature long-range magnetic ordering[11,12].

Remarkably, many micrometer-sized bimeron-like domains are observed in trilayer and another few-layer NiI$_2$ across the entire range of sweeping magnetic field[27]. The spin-up and spin-down domains exist in pairs (Fig. 2a and Extended Data Fig. 3). Two typical bimeron-like domains in trilayer NiI$_2$ at 0 T and 295 K are shown in Extended Data Fig. 4. The RMCD signals in each bimeron-like domain display opposite sign and nearly equal intensities. The magnetic moments point upwards or downwards in the core region and gradually decrease away from the core, and approaches zero near the perimeter. This magnetic moment distribution possibly indicates a pair of topological spin meron and antimeron with opposite chirality in a cycloid ground state[28,29]. The bimeron-like domains in both of the two few-layer NiI$_2$ shrink and weaken as the temperature increase from 10 K to 295 K (Extended Data Fig. 5), however the bimeron-like magnetization textures remain robust in all temperature, indicating the bimeron-like domains are thermodynamically stable. The high stability of the bimeron-like magnetic domains probably originate from the topological protection, which also contributes to the preservation of magnetization even if upon a reversal magnetic field of 1 T.

Figure 2b shows the temperature-dependent RMCD loops in a non-bimeron-like area of the trilayer $NiI_2$ sweeping between +3 T and -3 T. The RMCD loops show a highly nonlinear behavior with magnetic field[30] and plateau behaviors for the out-of-plane magnetization. At 10 K, the RMCD intensity near 0 T is suppressed and approaches zero, suggesting the vanishing remnant magnetization, which indicates a compensation of the out-of-plane magnetization and non-collinear AFM coupling in the trilayer $NiI_2$. And the gradual increases of the RMCD signal are observed with increasing magnetic field between ±1.2 and ±2.6 T, suggesting a spin-canting process. With increasing temperature, the plateaus between 1.2 T and -1.2 T is gradually weakened; and the nonlinear plateaus behavior nearly switches to a linear spin-canting behavior with magnetic field at 295 K. Similar magnetic hysteresis loops have been demonstrated in magnetic moiré superlattices[30] and skyrmions[31], which show definite non-collinear spin textures.

Figure 2c-d show the corresponding temperature dependence of RMCD intensities as a function of magnetic field. A phase transition takes place at around 35 K in the absence of magnetic field (Fig. 2c-d and Extended Data Fig. 6), validating that the as-exfoliated $NiI_2$ is trilayer[22,32]. The magnetic transition temperature ($T_{m1}$, $T_{m2}$) is field-dependent, and three magnetic phases are strikingly observed under positive field. Below 35 K and 0.5 T, the RMCD nearly decrease to zero and the non-collinear AFM coupling is predominated. With increasing temperature, a transition of magnetic phase takes place, in which the out-of-plane upward magnetizations significantly increase, implying a magnetic transition from non-collinear AFM states to pseudo-ferromagnetism states. The pseudo-ferromagnetism with nonzero upward magnetization is persisted to near room temperature, which is consistent with the RMCD maps (Extended Data Fig. 3-5). The unconventional weak pseudo-ferromagnetic induced by spin canting have also been observed in non-collinear antiferromagnet $Mn_3Sn$ crystals with kagome lattice at room temperature[33]. The transition temperature $T_{m1}$ decreases as the positive magnetic field increase, while slightly increases with increasing negative magnetic field (Extended Data Fig. 6a-b), indicating spontaneous upward spin canting with increasing temperature. The positive magnetic field favor the upward spin canting, and the magnetic phase transition takes place at lower temperature, but the negative magnetic field induce a spin-down energy barrier to block the spin-up canting, and thus a higher temperature is needed. Upon a higher magnetic field, another magnetic transition is observed. The transition temperature $T_{m2}$ also decreases as the positive magnetic field increase, while nearly keep constant with increasing negative magnetic field (Fig. 2d and Extended Data Fig. 6c), further suggesting spontaneous upward spin canting. Previous report has shown that the Curie temperature of $NiI_2$ crystal can be improved to 310 K by the pressure[34]. Recent work has demonstrate a room-temperature ferromagnetism of thin $NiI_2$ flakes due to decreasing surface symmetry, suggesting that few-layer $NiI_2$ drastically differs from bulk $NiI_2$[35]. But further deep studies must be done to reveal the exact physical mechanism.

## Ferroelectricity in trilayer $NiI_2$ device

To determine ferroelectricity in few-layer $NiI_2$ device, we performed the temperature- and frequency-dependent measurement of electric polarization via *I-E* and *P-E* hysteresis loops, which allows an accurate estimation of the ferroelectric polarization of few-layer flakes[36]. We fabricated two vertical heterostructure devices of graphene/hBN/$NiI_2$/graphene/hBN (Fig. 1a and Extended Data Fig. 1). The hBN flake was used as an excellent insulating layer to prevent large leakage current and guarantee the detections of ferroelectric (FE) features[36,37] (Extended Data Fig. 7). The hBN insulator shows a linear *P-E* behavior and a rectangle-shaped *I-E* loops (Extended Data Fig. 8), indicating excellent insulativity for ferroelectric hysteresis measurements (see Methods for details)[38,39]. The temperature and frequency-dependent $I$-$E$ and $P$-$E$ loops are shown in Fig. 3 and Extended Data Fig. 9-13, and the forward and backward scans of the electric polarization as a function of electric field show characteristic ferroelectric $I$-$E$ and $P$-$E$ hysteresis from 10 K to 295 K. Strikingly, a characteristic double-hysteresis loop of antiferroelectric (AFE) polarization emerges accompanied with decreasing remanent polarization ($P_r$). More importantly, a pair of opposite single peaks of switching current ($I$) are observed when sweeping voltage at 5 Hz, which is attribute to charge displacement and implies two stable states with inverse polarity (Fig. 3b and e). Whereas two pair of opposite bimodal peaks are observed when sweeping voltage at 1Hz, which is attribute to AFE-FE and FE-AFE transitions under electric field sweeping (Fig 3b and 3e )[40]. This suggests an evolution from FE to AFE polarization with frequency is observed[41,42], exhibiting the decisive evidence of ferroelectricity. As shown in Fig. 3a, the $P_r$ of ferroelectric polarization reaches to 0.33 nC/mm$^2$ at 10 K (Device 1, Top, Blue circle 1); as the frequency of electric field decrease, the $P_r$ decrease to 0.21 nC/mm$^2$ and the ferroelectric hysteresis loop become pinched arising from the coexistence of ferroelectric and antiferroelectric[43,44] (Middle, Blue circle 2), and ultimately, a distinct double-hysteresis loop with a weak $P_r$ of 0.10 nC/mm$^2$ occurs (Bottom, Blue circle 3), demonstrating an evolution from FE to AFE. Similar FE-AFE evolutions with frequency have been observed ranging from 10 K to 150 K (Fig. 3a, b, e and Extended Data Fig. 9). Remarkably, the double-hysteresis loop of AFE polarization become thinner along with sharp current peaks of FE-AFE transition, and an extremely slim double-hysteresis $P$-$E$ loop is observed at 100 K and 150 K (Bottom, Extended Data Fig. 9a and c), in which a linear polarization response to the electric field takes place between -0.4 to +0.4 MV/cm and two small hysteresis splitting only emerge above 0.4 MV/cm. This has been theoretically predicted in topological domain antiferroelectrics[45]. Moiré domain antiferroelectrics have been recently reported to show a near-zero remanent polarization due to topological protection at room temperature[43]. The topological antiferroelectric features might be counterparts of bimeron-like magnetic domains in the NiI$_2$ non-collinear magnets (Extended Data Fig. 4 and 5), associated by the magnetoelectric coupling effect.

The typical $I$-$E$ and $P$-$E$ loops of FE and AFE polarizations are still detected at 230 K and 270 K, and the evolutions from FE to AFE are also distinctly observed. One pair of switching current peaks in $I$-$E$ loops turns into two pairs with decreasing frequency, at the same time, the single-hysteresis $P$-$E$ loop is pinched and changed to a double-hysteresis loop (Extended Data Fig. 10 and 11). At 295 K, the $I$-$E$ loops still manifest the evolution of single to two pairs of opposite switching current peaks and the $P$-$E$ loops still show vivid ferroelectric hysteresis behaviors, although the FE features become weaker, in particular, after subtracting the hBN current background (Extended Data Fig. 8 and 12). To confirm the room-temperature ferroelectric again, another NiI$_2$ device were prepared (Device 2), which show evident $I$-$E$ and $P$-$E$ characteristic loops of FE and AFE at 295 K, typically with stronger AFE features (Fig. 3c-d and Extended Data Fig. 13). As decreasing frequency of electric field from 1 KHz to 0.5 KHz, two pairs of opposite switching current peaks are clearly observed (Fig 3d and 3f) and the $P_r$ decrease to near zero, leading to that the FE single loop evolves to characteristic AFE double loop. With further decreasing to 0.2 KHz, the current peaks of AFE-FE transition are more evident and the AFE double loop is thinned and slim. Moreover, the switching current peaks are broadened with increasing temperature, and around 5 times wider at 295 K than that at 10 K (Fig. 3e-f, Extended Data Fig. 12c and 14). From another point of view, this gives another evidence of the ferroelectric, suggesting that the increasing temperature shrinks the FE domain size and weakens the domain intercoupling, leading to broad current peaks and small hysteresis loops[46]. This comprehensive temperature- and frequency-dependent evolution behaviors confirm the FE and AFE behaviors in trilayer NiI$_2$ device across the entire range from 10 K to 295 K, and suggest a finding of a room-temperature 2D vdW multiferroic.

To further investigate the switching dynamics of ferroelectric polarization, we measured the $P$-$E$ loops as a function of frequency in the temperature range from 10 K to 295 K under the same electric field. Figure 3g illustrates the extracted temperature-dependent $P_r$ curves as a function of frequency. As the frequency increases, the $P_r^+$ ($P_r^-$) initially rises and reaches a maximum value at a certain frequency, which is defined as resonance frequency $f_R^+$ ($f_R^-$). The resonance peaks shift progressively toward the higher frequency with increasing temperature. In addition, the $f_R^-$ is significantly higher than $f_R^+$ at low temperature, and they tend to be equal at higher temperature. The anomalous behaviors of resonance frequency are consistent with the asymmetric magnetic phase transitions at opposite magnetic field (Fig. 2d and Extended Data Fig. 6), suggesting strong correlations between ferroelectric polarization and magnetism in the trilayer $NiI_2$. The $f_R^+$ ($f_R^-$) follows an Arrhenius relationship (it can also be explained as the nature of Merz's law mentioned below) and the plot is clearly divided into two sections at around 35 K (Fig. 3h), indicating that two thermal activation processes occurs in the $NiI_2$ trilayer. Remarkably, a transition of $P_r$ at resonance frequencies ($P_{r(max)}$) also occurs at around 35 K (Extended Data Fig. 15), which is consistent with the magnetic phase transition from RMCD and implies strong magnetoelectric coupling in the trilayer $NiI_2$. To carefully verify the physical meaning of resonance frequency in the trilayer $NiI_2$, we calculate the switching time via a classical KAI model (Part A, Extended Data Fig. 16 and 20)[47], where the reciprocal of switching time ($1/\tau$) follows an Arrhenius relationship with temperature, and also divided into two sections at around 35 K (Fig. 3h), consistent with the temperature-dependent behavior of $f_R$, which suggests that the reciprocal of the resonance frequencies ($1/f_R$) is the average switching time ($\tau$) of ferroelectric domains. Furthermore, the switching dynamics of domain obeys generally recognized Merz's law over the 230-295 K temperature range (Fig. 3i), which is consistent with the two-dimensional growth of critical nuclei via molecular dynamics simulations[48]. The inset in Fig. 3i shows the temperature dependence of the activation field $E_{a,t} = UE_{C0}/(k_BT)$ above 230 K. The nearly linear relationship between $E_{a,t}$ and $1/T$ gives a temperature-independent constant of ~810 K MV/cm, which is consistent with intrinsic ferroelectric switching mechanism[49].

**Mutual control of magnetism and ferroelectricity**

To reveal the magnetoelectric coupling effect, we studied the mutual control of magnetic and ferroelectric properties in the trilayer $NiI_2$ device. Figure 4a shows the spatial RMCD maps under different voltage at 295 K, collected at zero magnetic field. As the electric field is increased from 0 to +2.22 MV/cm (6 V), the upward ferroelectric polarization is gradually enhanced, the upward magnetization intensities are also progressively enhanced, meanwhile the upward magnetization areas are significantly enlarged, which nearly extend to the entire trilayer $NiI_2$ at +1.85 MV/cm (5 V), and magnetization is nearly unchanged at higher electric field of +2.22 MV/cm. This is because the ferroelectric polarization is gradually saturated when the electric field reaches about 1.5 MV/cm. When the electric field is reversed to negative value ranging from -0.37 MV/cm (-1 V) to -2.22 MV/cm (-6 V), similar manipulation of magnetic textures is observed (Extended Data Fig. 17). The RMCD intensities of typical individual magnetic domains and the sum of each RMCD intensity extracted from the spatial map are plotted as a function of electric field, which signifies an even symmetry (Fig. 4b). It is shown that upward magnetization increases with increasing electric field and reach saturation at an electric field of 1.85 MV/cm, indicating that the electric control of magnetism arising from the intrinsic coupling of ferroelectric and magnetism. Alternatively, the magnetic and electrical control of bimeron-like domains at 295 K is shown in Extended Data Fig. 18a-c. As the electric field is increased from 0 to 2.22 MV/cm (6 V), the out-of-plane components of both "core-down antivortex" (blue area) and "core-up vortex" (red area) are gradually enhanced (Extended Data Fig. 18a). With the increasing negative electric field (Extended Data Fig. 18b), the "core-down antivortex" and "core-up vortex" show similar evolution to that under positive electric field, which indicates that the magnetization is coupled with electric polarization, but independent of the direction of electric polarization, and shows an even symmetry with electric field. However, the magnetic control of bimeron-like domains shows odd symmetry and entirely differs from the electrical control, where the downward (upward) magnetization normally increases with increasing downward (upward) magnetic field (Extended Data Fig. 18c). This observation of magnetization enhancement and magnetic texture expansion suggests an efficient and robust control of magnetism by electric field under zero magnetic field at 295 K, pointing to the potential use of few-layer $NiI_2$ as a low-energy magnetoelectric memory.

Furthermore, we study the magnetic control of ferroelectric properties, as shown in Fig. 4c-f. The $P_r$ extracted from the $P$-$E$ hysteresis loop is plotted as a function of frequency at different magnetic field in the range from 10 K to 295 K (Fig. 4c and Extended Data Fig. 19). The resonance peaks are unambiguously observed and the resonance frequency is lowered with increasing magnetic field, which indicate that the magnetic field reduces the velocity of ferroelectric domain switching, i.e. "magnetic field freezing effect". To better understand the magnetic control behavior, the switching time of ferroelectric domain under different magnetic fields in the range of 10-295 K is calculated by KAI model (Fig. 4d and Extended Data Fig. 20 and Part B). The switching time $\tau$ increase as magnetic field increase, which also signifies an even symmetry, consistent with the electrical control of magnetization as shown in Fig. 4e. At 10 K, the switching time $\tau$ represents a plateaus behavior between around ±1.5 T, and with increasing temperature, the plateaus-like behavior is gradually weakened and is persisted to 60 K, which is consistent with the evolution of the RCMD plateaus with temperature as shown in Fig. 2b. Figure 4f shows the temperature- and magnetic-field dependence of magnetoelectric coupling obtained from Fig. 4e. When the temperature increases from 10 to 295 K, the magnetoelectric coupling gradually becomes robust, leading to a maximum enhancement of switching time by 49% (-7 T) in the vicinity of 150 K. At 295 K, the increase of switching time reaches 31%.

In summary, we report a 2D vdW multiferroic trilayer $NiI_2$. We observed strong evidences for the coexistence of ferroelectric and magnetic order ranging from 10 K to 295 K via scanning RMCD maps, $P$-$E$ and $I$-$E$ hysteresis loop. Near room temperature, we achieve unprecedented mutual control of ferroelectric and magnetic properties in the $NiI_2$ trilayer device. We envision that the near-room-temperature 2D vdW multiferroic $NiI_2$ will provide numerous opportunities for exploring fundamental low-dimensional physics, and for creating faster, smaller, lower energy memory devices, and will introduce a paradigm shift for engineering new ultra-compact spintronics and brain-inspired chip.

**Online content**

Any methods, additional references, Nature Portfolio reporting summaries, source data, extended data, supplementary information, acknowledgements, peer review information; details of author contributions and competing interests; and statements of data availability are available at https:

## Methods

### Sample fabrication

$NiI_2$ flakes were mechanically exfoliated from bulk crystals via PDMS films in a glovebox, which were synthesized by chemical vapor transport method from elemental precursors with molar ratio Ni:I = 1:2. All exfoliated hBN, $NiI_2$ and graphene flakes were transferred onto pre-patterned Au electrodes on $SiO_2$/Si substrates one by one to create heterostructure in glovebox, which were further in-situ loaded into a microscopy optical cryostat for magneto-optical-electric joint-measurement. The whole process of $NiI_2$ sample fabrications and magneto-optical-electric measurements were kept out of atmosphere.

### Magneto-optical-electric joint-measurement

The polar RMCD, white-light MCD, Raman measurements and ferroelectric *P-E* and *I-E* measurements were performed on a powerful magneto-optical-electric joint-measurement scanning imaging system (MOEJSI)[19], with a spatial resolution reaching diffraction limit. The MOEJSI system was built based on a Witec Alpha 300R Plus low-wavenumber confocal Raman microscope, integrated with a closed cycle superconducting magnet (7 T) with a room temperature bore and a closed cycle cryogen-free microscopy optical cryostat (10 K) with a specially designed snout sample mount and electronic transport measurement assemblies.

The Raman signals were recorded by the Witec Alpha 300R Plus low-wavenumber confocal Raman microscope system, including a spectrometer (150, 600 and 1800/mm) and a TE-cooling Andor CCD. A 532 nm laser of ~0.2 mW is parallel to the X-axis (0º) and focused onto samples by a long working distance 50× objective (NA = 0.55, Zeiss) after passing through a quarter-wave plate (1/4λ). The circular polarization resolved Raman signals passed through the same 1/4λ waveplate and a linear polarizer, obtained by the spectrometer (1800/mm) and the CCD.

For white-light MCD measurements, white light with Köhler illumination from Witec Alpha 300R Plus microscope was linearly polarized at 0º by a visible wire grid polarizer, passed through an achromatic quarter-wave (1/4λ) plate and focused onto samples by a long working distance 50× objective (Zeiss, NA = 0.55). The right-handed and left-handed circularly polarized white light was obtained by rotating 1/4λ waveplate at +45º and -45º. The white-light spectra were recorded by the Witec Alpha 300R Plus confocal Raman microscope system (spectrometer, 150/mm). The absorption spectra of right-handed and left-handed circularly polarized light in different magnetic field can be obtained as the previous work[25], giving corresponding MCD spectra.

For polar RMCD measurements, a free-space 532 nm laser (2.33 eV) of ~2 μW modulated by photoelastic modulator (PEM, 50 KHz) was reflected by a non-polarizing beamsplitter cube (R/T = 30/70) and then directly focused onto samples by a long working distance 50× objective (NA = 0.55, Zeiss), with a diffraction limit spatial resolution of ~590 nm. The reflected beam which was collected by the same objective passed through the same non-polarizing beamsplitter cube and was detected by a photomultiplier (PMT), which was coupled with lock-in amplifier, Witec scanning imaging system, superconducting magnet, voltage source meter and ferroelectric tester.

Ferroelectric *P-E* and *I-E* hysteresis loop of a $NiI_2$ device of Gr/hBN/$NiI_2$/Gr were measured by classical ferroelectric measurements and directly recorded by ferroelectric tester (Precision Premier II: Hysteresis measurement), which were contacted with the top and bottom graphene electrodes by patterned Au electrodes (Fig. 1a) through the electronic assemblies of the microscopy optical cryostat. The mechanism of ferroelectric measurement has been given by previous work[50]. The detected signals include two components: a ferroelectric term of NiI$_2$ ($2P_rA$) and a linear non-ferroelectric term of hBN insulator ($\sigma EAt$), $Q = Q_{NiI} + Q_{BN} = 2P_rA + \sigma EAt$. If only hBN insulator, a linear *P-E* loop take place, consistent with our experimental results of hBN flake (Extended Data Fig. 8). The linear hBN background have no effect on the ferroelectric features, and hBN flakes as excellent insulator suppress and overcome the leakage current, which for guarantee the detections of NiI$_2$ ferroelectric features[36-39].

**STEM Imaging, Processing, and Simulation**

Atomic-resolution ADF-STEM imaging was performed on an aberration-corrected JEOL ARM 200F microscope equipped with a cold field-emission gun operating at 80 kV. The convergence semiangle of the probe was around 30 mrad. Image simulations were performed with the Prismatic package, assuming an aberration-free probe with a probe size of approximately 1 Å. The convergence semiangle and accelerating voltage were in line with the experiments. The collection angle for ADF imaging was between 81 and 228 mrad. ADF-STEM images were filtered by Gaussian filters, and the positions of atomic columns were located by finding the local maxima of the filtered series.

**Acknowledgments**

B.P. and L.D. acknowledge support from National Science Foundation of China (52021001). B.P. acknowledge support from National Science Foundation of China (62250073). R.C.C. acknowledge support from National Science Foundation of China (52231007). H.L. acknowledge support from National Science Foundation of China (51972046). L.D. acknowledge support from Sichuan Provincial Science and Technology Department (Grant No. 99203070). L.D. acknowledge support from Sichuan Provincial Science and Technology Department (Grant No. 99203070). L.Q. acknowledge support from National Science Foundation of China (520720591 and 11774044). J.W. thanks the National Natural Science Foundation of China (Grant No. 11974422), the Strategic Priority Research Program of the Chinese Academy of Sciences (Grant No. XDB30000000).

**Author contributions**

B.P conceived the project. Y.W. prepared the samples and performed the magneto-optical-electric joint-measurements and Raman measurements assisted by B.P., and performed the ferroelectric measurements assisted by L.Q., and analyzed and interpreted the results assisted by H.L., N. L., W.J., L.D. and B.P.. C. Y, R.C, X.X. and X.H. performed the STEM measurements. Y.W. and B.P. wrote the paper with input from all authors. All authors discussed the results.

**Competing interests**

The authors declare no competing interests.

**Additional information**

**Supplementary information** The online version contains supplementary material available at

https:
**Correspondence and requests for materials** should be addressed to Bo Peng.
**Reprints and permissions information** is available at www.nature.com/reprints.

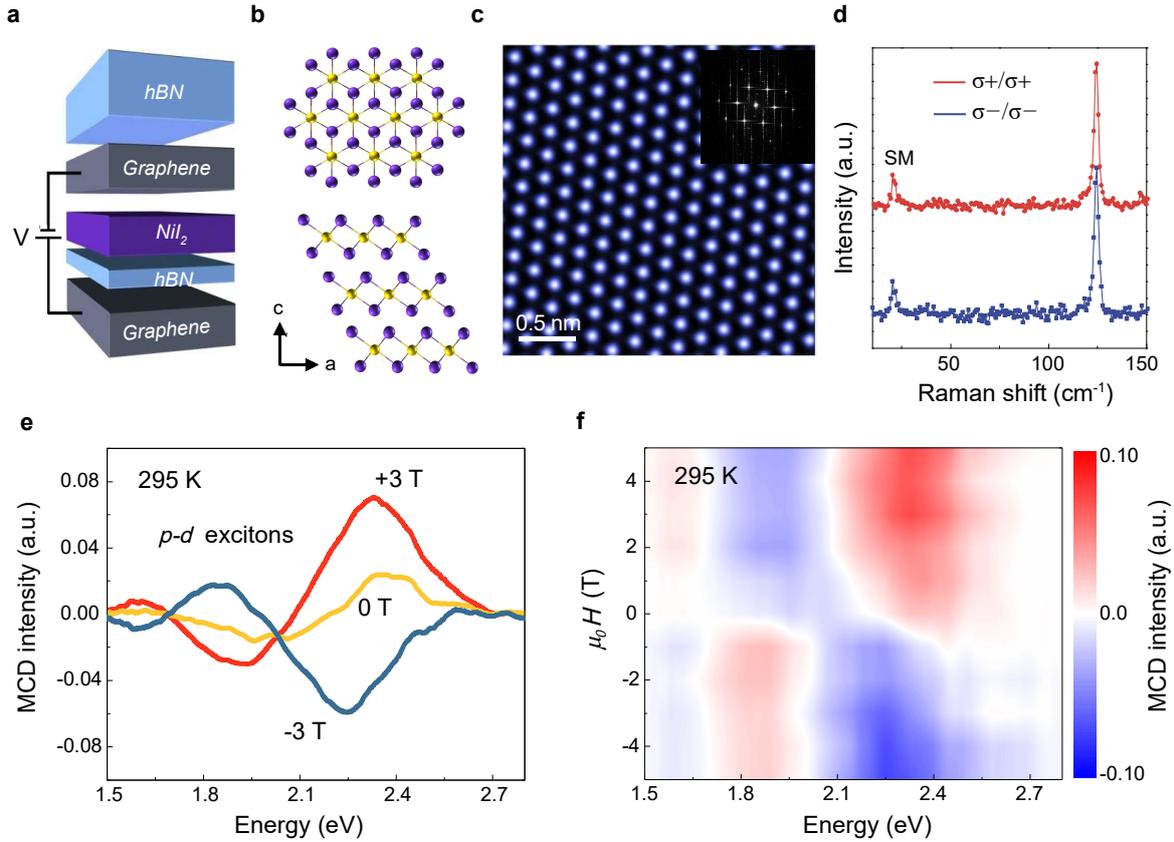

**Fig. 1 | Crystal structure, MCD measurements of trilayer NiI$_2$. a,** Schematic of trilayer NiI$_2$ sandwiched between graphene and hBN. **b,** View of the in-plane and out-of-plane atomic lattice. The magnetic Ni$^{2+}$ ions are surrounded by the octahedron of I$^-$ ions, and three NiI$_2$ layers as a repeating unit stack in a staggered fashion along the *c* axis. **c,** Atomic-resolution ADF-STEM image showing signature hexagonal patterns of rhombohedral stacking in few-layer NiI$_2$ crystals. The inset shows the corresponding FFT image. **d,** Circular polarization resolved Raman spectra of a trilayer NiI$_2$ device (Fig. 1a) at 295 K, excited by 532 nm laser. "SM" indicates the interlayer shear mode of trilayer NiI$_2$. **e,** Smoothed white-light MCD spectra of trilayer NiI$_2$ at +3 T, 0 T and -3T at 295 K. **f,** Corresponding white-light MCD color map versus magnetic field and wavelength at 295 K. MCD signals are sensitive to spin electronic transitions and magnetic moments in the electronic states. The MCD features are spin-sign dependent and reverse as magnetic field switch. The non-zero remanent MCD signals at ~2.3 eV at 0 T suggest long-range magnetic orders, inducing an asymmetric magnetic-field dependence.

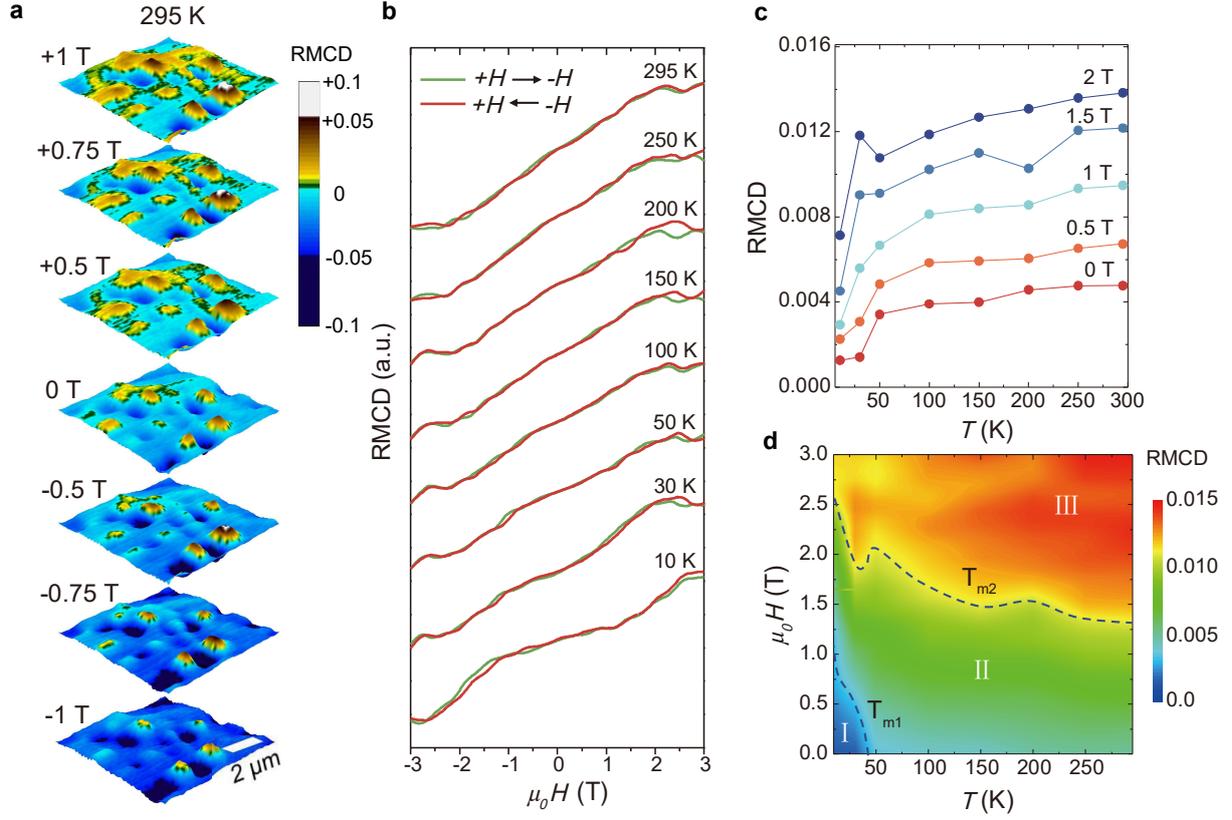

**Fig. 2 | Magnetism in trilayer NiI$_2$ device.** **a,** Polar RMCD maps upon a 2.33 eV laser with diffraction-limited spatial resolution (see Methods), collected at 295 K and selected magnetic field. Magnetic domains can be reversed through switching magnetic field, validating near-room-temperature magnetism[11,12]. **b,** Temperature-dependent RMCD curves sweeping between +3 T and -3 T from 10 K to 295 K, suggesting a non-collinear spin texture[30], and inducing the emergence of non-zero magnetization between ±1.5 T from 10 K to 295 K. **c,** The RMCD intensities as a function of temperature at selected magnetic field. **d,** Magnetic field-temperature ($\mu_0H$-$T$) magnetic phase diagram, based on the magnetic-field and temperature dependence of RMCD intensities. With magnetic field and temperature, three magnetic phases marked with I, II and III are observed. The magnetic phase temperature ($T_{m1}$, $T_{m2}$) decrease with magnetic field, originating from spontaneous upward spin canting with increasing temperature. The dashed lines are guidelines for the eyes, indicating the phase transition temperature of $T_{m1}$ and $T_{m2}$.

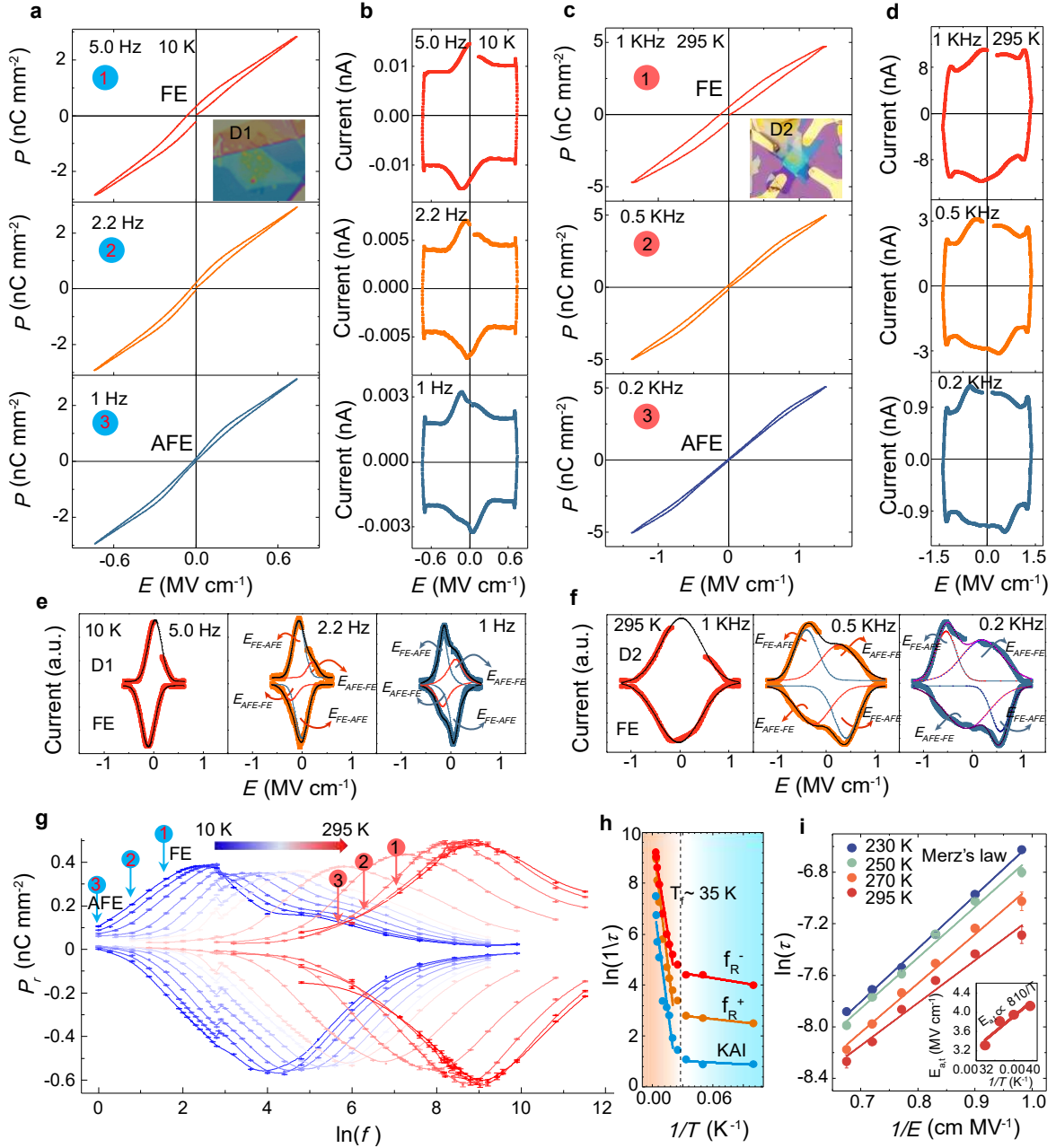

**Fig. 3 | Existence of ferroelectric and anti-ferroelectric orders in trilayer NiI$_2$ device. a-d,** Raw *P-E* and *I-E* loops at various frequencies from device 1 (D1) and 2 (D2), resembling the behaviors of relaxor ferroelectrics. The *P-E* and *I-E* loops were tested by classical ferroelectric measurement methods and directly recorded by a ferroelectric tester through the top and bottom graphene (Fig. 1a, S1). **e, f,** Corresponding *I-E* loops from Fig. 3b and 3d (bottom) subtracted the current background. Two pairs of current peaks (FE-AFE and AFE-FE switching peaks) were obtained by Gaussian fitting. An evolution from FE to AFE was observed. **g,** Temperature-dependent remanent polarization ($P_r$) curves *vs* frequency (Hz). The error bars are standard deviations of $P_r$. The frequency with maximum $P_r^+$ ($P_r^-$) is simply defined as resonance frequency $f_R^+$ ($f_R^-$). **h,** The $f_R^+$, $f_R^-$, and the reciprocal of switching time ($1/\tau$) by KAI model *vs* temperatures at 0 T. The lines are fits according to Arrhenius relationship. **i,** Plot of $\ln(\tau)$ versus $1/E$ curves at different temperatures. The inset shows the temperature dependence of the activation field $E_{a,t}$.

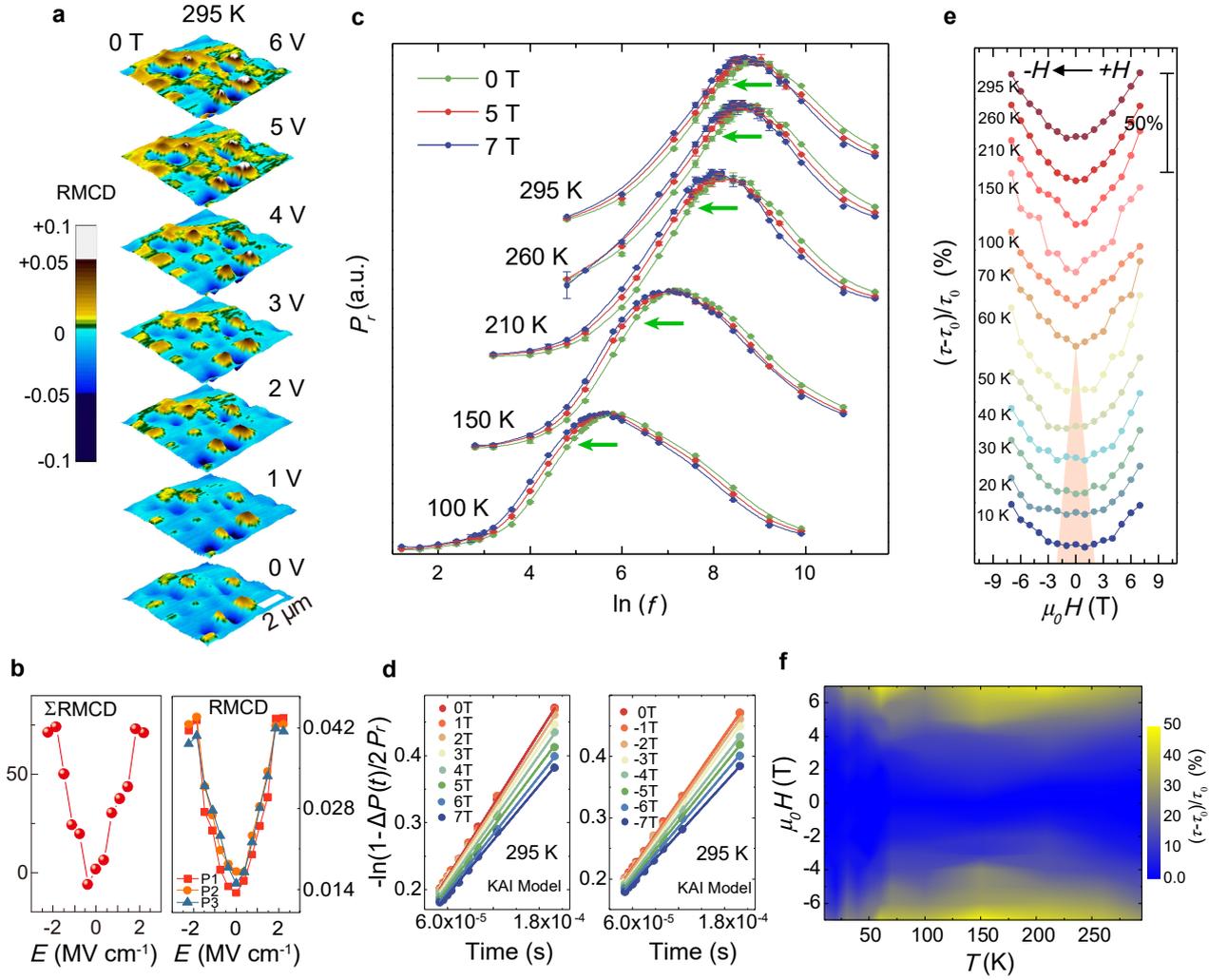

**Fig. 4 | Mutual control of magnetism and ferroelectricity in trilayer NiI₂ device**. **a,** Polar RMCD maps upon applied voltage ranging from 0 V to 6 V (2.22 MV/cm) at 295 K, collected at 0 T. Electrical control of magnetic textures is clearly observed in the trilayer NiI₂ at 295 K. **b,** Total RMCD intensities as a function of electric field, extracted from Fig. 4a, and the RMCD intensity of typical three individual magnetic domains with electric field (as labeled in Extended Data Fig. 17). **c,** The frequency-dependent remanent polarization $P_r$ curves in 0, 5 and 7 T magnetic field at various temperatures. The error bars are standard deviations of $P_r$. The frequencies decrease with increasing magnetic field, implying a magnetic-field freezing effect. **d,** Fitting by KAI model for different magnetic field at 295 K, giving the switching time τ. **e,** The $(\tau-\tau_0)/\tau_0$ as a function of magnetic field at various temperatures, indicating a degree of magnetic control of switching time, where τ and $\tau_0$ is switching time in a magnetic field and without magnetic field, respectively. The magnetic field sweeps from positive to negative. The orange shaded area highlights the evolution of the plateaus. **f,** The color map of $(\tau-\tau_0)/\tau_0$ as a function of magnetic field and temperature. All magnetic control experiments were carried out under the same electric field.

15